\documentstyle[preprint,aps,pra,epsf]{revtex}

\def\alli{Al$_n$Li$_n$}            %  Macro  for  Al_n Li_n
\def\two{Al$_2$Li$_2$}             %  Macro  for  n=2
\def\three{Al$_3$Li$_3$}           %  Macro  for  n=3
\def\four{Al$_4$Li$_4$}            %  Macro  for  n=4
\def\five{Al$_5$Li$_5$}            %  Macro  for  n=5
\def\six{Al$_6$Li$_6$}             %  Macro  for  n=6
\def\seven{Al$_7$Li$_7$}           %  Macro  for  n=7
\def\eight{Al$_8$Li$_8$}           %  Macro  for  n=8
\def\nine{Al$_9$Li$_9$}            %  Macro  for  n=9
\def\ten{Al$_{10}$Li$_{10}$}       %  Macro  for  n=10
\def\eleven{Al$_{11}$Li$_{11}$}    %  Macro  for  n=11
\def\alsix{Al$_6$Li$_8$}           %  Macro  for  Al_6-Li_8

\def\et{{\em et al}}               %  Macro  for  {\em et al}

\title {
          Evolution of the structural and bonding
          properties of Aluminum-Lithium clusters
       }

\author  { S. Chacko~\cite{email-chacko}, D. G. Kanhere~\cite{email-kanhere} }
\address {
           Department of Physics,
           University of Pune,
           Pune 411 007,
           India.
         }
\author  {
           V. V. Paranjape,
         }
\address {
           Department of Physics,
           Lakehead University,
           Thunder Bay,
           Ontario,
           P7B 5E1 Canada.
         }

\date{\today}

\begin{document}

\maketitle

\begin{abstract}

We present a systematic study of the geometries,
energetics, electronic structure and bonding in various
Al-Li clusters viz. \alli ~($n$=1-11), Al$_2^-$,
Al$_2^{2-}$, Al$_2$Li, Al$_2$Li$^-$, and \alsix ~using
Born-Oppenheimer molecular dynamics within the
framework of density functional theory.
The growth pattern in these clusters is divided in
two broad categories: the first one consisting of a
bent rhombus of \two ~($n$=2-4) and the second
one consisting of a pentagonal ring ($n$=7-9,11).
A substantial charge transfer is seen in nearly all
clusters with the exception of \two ~where the charge
transfer is not significant.
In clusters with more than six Al atoms, the
eigenvalue spectrum is divided into two groups: a lower
group of jellium-like states and a higher group of
localized bonds
formed through the interaction of the
$p$ electrons on each of the Al atoms.
Finally, we have discussed the tetravalent behavior of
Al atoms arising due to a charge transfer from the Li
atoms to the Al atoms.

\end{abstract}


\section{Introduction}
\label{intro}

The discovery of C$_{60}$
buckminsterfullerene~\cite{kroto1}, observation of
magicity in metallic clusters~\cite{knight1}, unusual
thermodynamical properties of clusters of atoms like
Na~\cite{haberland1}, Sn~\cite{jarold1} etc., has
given rise to an explosive growth of research in the
field of cluster physics.
Extensive theoretical and
experimental~\cite{book1,book2,parreviews} work has been
carried out to understand the physical and chemical
properties of clusters like size-evolutionary pattern
in cluster geometry, thermodynamics, vibrational and
rotational properties, optical properties, electronic
structure, and bonding, chemical reactivity, as well as
the emergence of the bulk properties with the increase
in cluster size.

Recently observed aromaticity~\cite{wang1} and
antiaromaticity~\cite{sharan1,sharan2,sharan3,wang2} in
inorganic molecules, and in particular Al-Li clusters,
has stimulated further research in these systems.
These properties are well-known and important in
organic chemistry~\cite{aromaticity1}.
However, recently the first experimental and
theoretical evidence of aromaticity in all-metal system
viz. Al$_4^{2-}$ and $M$Al$_4^-$ ($M$ = Li, Na, Cu) was
reported by Wang and coworkers~\cite{wang1}.
They found the Al$_4^{2-}$ dianion to be square-planar
possessing two $\pi$-electrons, thus conforming the
structural criterion for aromaticity.
The concept of antiaromaticity, introduced by
Breslow~\cite{breslow1} \et., corresponds to the
destabilization seen in the cyclic systems with
$4n~\pi$-electrons.
Although such molecules have not yet been observed
except in organic chemistry, calculations reported by
Shetty~\cite{sharan1,sharan3} \et. and
others~\cite{wang2} indicate the \four ~cluster to be
antiaromatic.

Aluminum-lithium clusters, which we study in this
paper, are also interesting due to their unusual
structural and bonding properties that are quite
different from those of the pure clusters of the
constituent elements.
Both pure Al and Li clusters have been extensively
studied using the spherical jellium model
(SJM)~\cite{knight1}.
However, it is observed that although the SJM has
been quite successful in describing the gross
electronic structure and the stability of alkali metal
atom clusters~\cite{knight1,liclus}, the Al clusters
present an interesting contrast.
Rao and Jena~\cite{rao1} have reported a comprehensive
study of the aluminum cluster using {\em ab initio}
density functional theory.
They found that in small Al clusters, the effective
valency of Al is one~\cite{rao1} due to a large energy
gap of about 5eV separating the $3s^2$ and $3p^1$
orbitals.
However, Rao~\cite{rao2} \et. as well as
Dhavale~\cite{dhavale1} \et. did not find any signature
of such monovalent behavior, contrary to the
expectations.
Investigation by Cheng~\cite{cheng1} \et. shows that a
single Al atom in Li clusters introduces a localized
bond between the Al impurity and the Li
host.
They also found a magic cluster AlLi$_5$, and suggested
that Al$_n$Li$_{5n}$ can aggregate at least for some
values of $n\ge2$ and may exhibit properties
characteristic to an assembly of such clusters from
the AlLi$_5$ subunits.
Akola~\cite{akola1} \et., however, found that this idea
does not apply after $n=2$.
Further, their investigation on bonding in Li-rich
Al-Li clusters show a substantial charge transfer from
Li atom and nearby Al atom, strengthening the ionic
Al-Li bond, while the Al-Al bond gained a more covalent
nature.
Such a charge transfer has been observed by
Kumar~\cite{kumar1} in several mixed Al-Li clusters.
He found a layered Al$_{10}$Li$_8$ compound to
be magic with electronic and geometric shell closing.
This was the first instance where a shell closure in
$s-p$ bonded metal cluster was found to occur at 38
valence electrons.
Note that the magic numbers for alkali metal atom
clusters~\cite{knight1} are 2, 8, 18, 20, 40,...,
whereas, some of the magic Al clusters are Al$_7^+$,
Al$_7^-$, Al$_{11}^-$, Al$_{13}^-$,
etc~\cite{rao1,li1}.

The Al-Li bulk is especially interesting due to the
fact that it forms a stable alloy at over the wide
range of concentration.
However, the most stable intermetallic B32 phase
corresponds to the 50:50 concentration~\cite{freeman1}.
In this B32 phase, a mixture of covalent and ionic
bonding is seen due to a remarkable charge transfer
from the Li atoms to the Al atoms.
Moreover, the close resemblance of the density of
states (DOS) (without any band gap)~\cite{das1} to that
of the covalently bonded diamond structure indicates a
tetravalent behavior of Al atom.
Thus, based on these observations, one can expect that
the behavior of Al in \alli ~clusters would be similar
to that of the tetravalent atom Si~\cite{siclus},
Ge~\cite{geclus}, Sn, and Pb~\cite{bingwang1}.
Indeed, our present work shows that in certain
clusters, structure and bonding of Al$_n$ is similar to
that of the group IV~A clusters.
Moreover, earlier work on the heterogeneous Al-Li
clusters has focussed on specific aspects like
systematics in the geometry, stability, shell closure,
magic behavior etc.
Bonding in such clusters was discussed on the basis of
total charge density as well as the difference of the
self-consistent charge density, $\rho_{scf}$, and the
superimposed atomic charge densities
$\rho_{superimposed}$ of the constituent atoms.
On the contrary, analysis of the molecular orbitals
(MOs) has revealed some unusual properties like
aromaticity and antiaromaticity in all-metal Al-Li
clusters.
In our earlier work, we have examined some of the
issues concerning the geometry and the stability of
various Al-Li clusters viz. \alli ~($n$ =, 1-10,
13)~\cite{shah1}, Al$_n$Li$_7$ ($n$=1-7)~\cite{shah2},
AlLi$_n$ ($n$=1,8)~\cite{shah2} and Al$_{13}$Li$_n$
($n$=1-4,10,19,20,21)~\cite{abhijat1}.
We have found that the geometries of these clusters
were dictated by the geometry of the core Al cluster,
enclosed in the Li cage.
However, since these calculations were performed by the
density based molecular dynamics (DBMD), an analyses of
the bonding and electronic structure could not be
done.
In the present work, we study the systematics of the
geometries, energetics, electronic structure and the
bonding properties in various Al-Li clusters viz. \alli
~($n$=1-11), Al$_2^-$, Al$_2^{2-}$, Al$_2$Li,
Al$_2$Li$^-$, and \alsix ~using Kohn-Sham formulation
of the density functional theory (DFT) within the
pseudopotential and the generalized gradient
approximation~(GGA).
The bonding in these clusters is analyzed via the
electron localization function (ELF)~\cite{silvi1} and
the molecular orbitals.
In section~\ref{numerics} we shall describe the
computational details, followed by a discussion of
the results in section~\ref{results}.

\section {Computational Details}

\label{numerics}

The ground state geometries as well as other low-lying
structures were obtained in two stages.
In the first stage, Born-Oppenheimer molecular dynamics
(BOMD)~\cite{payne1} based on Kohn-Sham (KS)~\cite{kohn1}
formulation of density functional theory (DFT) was
employed.
The computer code used for this purpose was developed
in our own group.
The total energy during each of the molecular dynamics
step was minimized using damped equation of
motion~\cite{joannopoulos1}.
The calculations were performed using norm-conserving
pseudopotentials of Bachelet~\cite{bachelet1} \et., in
the Kleinman-Bylander~\cite{kleinman1} form with
$s$~part treated as nonlocal.
The exchange-correlation potential was calculated using
the local density approximation (LDA) given by
Ceperley-Alder~\cite{ceperley1}.
A cubic supercell of length $40a.u.$ with an energy
cutoff of $\approx 17~Rydberg$ was found to provide
sufficient convergence of the total energy.
Increasing the energy cutoff did not improve the total
electronic energy significantly.

The damped equation of motion
scheme~\cite{joannopoulos1} permits use of a fairly
moderate time step $\approx$100a.u.  Starting from a
random configuration, the \alli ~clusters were heated
to 1400-1700$K$, and allowed to span the phase space
for a few thousand iterations.
In our previous
investigations~\cite{shah1,shah2,abhijat1} on \alli
~clusters, it was found that the Li atoms segregate at
the surface with the Al atoms forming an inner core.
In order to avoid any such segregation during the
molecular dynamics run, we have interchanged the Al and
the Li atoms.
This has ensured that the cluster visits all its local
minima as well as the global minimum.
At the end of each ionic displacement, the norm of the
eigenstates defined as $\mid
h{\psi}_i-{\epsilon}_i{\psi}_i \mid ^2$ (where
$\epsilon_i$ is an eigenvalue corresponding to the
eigenstate $\psi_i$ of the Kohn-Sham Hamiltonian $h$)
was within the range of $10^{-4}-10^{-7}$a.u.

In the second stage, various low-lying structures were
obtained by conjugate gradient and/or steepest
descent~\cite{payne1} method starting from various
suitable configurations during the molecular dynamics
run.
These configurations were then optimized using
ultrasoft pseudopotentials~\cite{vanderbilt1} within
the generalized gradient approximation (GGA)
implemented in the VASP~\cite{vasp} package.
The Perdew-Wang~\cite{perdew1} exchange-correlation
potential for GGA has been used.
The size of the simulation cell was varied according to
the cluster studied (see Table-I).

\begin{center}
Table - I
\vskip 0.5cm

Size of the supercell (in \AA) for various clusters,
where $n$ is the total number of atoms in the clusters.
\vskip 0.5cm

\begin{tabular} {|lllcl|}
\hline
& & & & \hspace{1.0cm} \\
\hspace{0.5in} & $n$ & \hspace{0.5in} ~ \hspace{0.5in} & Simulation Cell (\AA) &\\
& & & &\\
\hline
& & & &\\
&   2,4,6                &  ~  &  16x16x16  &\\
&   8,10,12              &  ~  &  18x18x18  &\\
&   14,16.18,20           &  ~  &  20x20x20  &\\
&& & &\\
\hline
\end{tabular}
\end{center}

The geometries were optimized with a
kinetic energy cutoff of the order of $\approx$
12~$Rydbergs$.
This energy cutoff was sufficient for the convergence
of the total energy.
Increasing the energy cutoff did not improve the total
electronic energy significantly.
The structures were considered to have converged when
the forces on each ion was less than 0.01eV/\AA ~with a
convergence in the total energy within the range of
$10^{-4}-10^{-6}$eV.
In general, we find that there are many isomeric
structures nearly degenerate to the lowest energy
state.
Few of the structures can be obtained by interchanging
the Al and the Li atoms or by rearranging the position
of Li atoms.
In the present work, we discuss only a few geometries
corresponding to the lowest energies.

The nature of the bonding has been investigated using
the electron localization function (ELF)~\cite{silvi1}
along with the molecular orbitals (MO) as well as the
difference charge density.
The ELFs have been found to be useful for elucidating
the bonding characteristics of a variety of systems,
especially in conjunction with the charge density.
The value of the ELF lies between 0 and 1, where 1
represents a perfect localization of the valence charge
density.
The difference charge density is the difference between
the self-consistent charge density, $\rho_{scf}$, and
the superimposed atomic charge densities,
$\rho_{superimposed}$, of the constituent atoms.

\section{Results and Discussion}
\label{results}

\subsection{Structure, energetics and stability}
\label{structure}

The ground state geometries and some of the low energy
structures of \alli, and \alsix ~clusters are shown in
figures~\ref{fig.str1} ($n$=2-6, and \alsix), and
\ref{fig.str2} ($n$=7-11).
The ground state geometries are divided into two broad
categories, for (i) $n$=2-4
(figures~\ref{fig.str1}a(i)-\ref{fig.str1}c(i)), and
(ii) $n$=7-9, and 11
(figures~\ref{fig.str2}a(i)-\ref{fig.str2}c(i),
\ref{fig.str2}e(i)), depending on the growth pattern.
In the first category a quinted roof-like structure of
\two ~is seen (a quinted roof structure is a rhombus
with a bend at its diagonal).
The second category shows a pentagonal ring which is a
signature of an icosahedral growth.
The ground state geometries of $n$=5, 6, and 10
clusters (figures~\ref{fig.str1}d, \ref{fig.str1}e,
\ref{fig.str2}d)  show neither a quinted roof nor a
pentagonal ring.
In what follows, we discuss the structure and stability
of these clusters.
The stability is discussed via the binding energy
(E$_b$), the dissociation energy ($\Delta$E) and the
second difference in the total energy with respect to a
single Al-Li pair ($\Delta^2$E), and the energy gap
between the highest occupied molecular orbital (HOMO)
and the lowest unoccupied molecular orbital (LUMO).
These quantities are plotted in
figure~\ref{fig.energetics}a (E$_b$),
\ref{fig.energetics}b ($\Delta$E, $\Delta^2$E), and
\ref{fig.energetics}c (HOMO-LUMO gap) as a function of
the number of Al atoms in the cluster and are defined
as

\begin{eqnarray}
\nonumber
{\rm E}_b[{\rm Al}_n{\rm Li}_n]
          = \Big
            [ -{\rm E}[{\rm Al}_n{\rm Li}_n]
          + n \left ( {\rm E}[{\rm Al}]
          + {\rm E}[{\rm Li}] \right ) \Big ] \Big / 2n\\
\nonumber
\Delta {\rm E}[{\rm Al}_n{\rm Li}_n]
          = \Big
            [ {\rm E} [{\rm Al}_n{\rm Li}_n]
          - {\rm E} [{\rm Al}_{n-1}{\rm Li}_{n-1}]
            \Big ]
          - {\rm E} [{\rm Al~Li}] \\
\nonumber
\Delta^2 {\rm E}[{\rm Al}_n{\rm Li}_n]
          = \Big [
            {\rm E} [{\rm Al}_{n-1}{\rm Li}_{n-1}]
          - 2 {\rm E} [{\rm Al}_n{\rm Li}_n]
            \Big ]
          - {\rm E} [{\rm Al}_{n+1}{\rm Li}_{n+1}]
\end{eqnarray}

\noindent
Let us recall that a pronounced maximum in the
dissociation energy $\Delta$E, along with a
corresponding minimum in $\Delta^2$E, signifies
stability.
In addition, in figure~\ref{fig.energetics}d, we have
plotted the smallest bond distances of Al-Al, Al-Li and
Li-Li as a function of the number of Al atoms in the
cluster.

The ground state geometry of \two ~is a bent rhombus
(figure~\ref{fig.str1}a(i)).
The planar geometry (figure~\ref{fig.str1}a(ii)) is the
low-lying structure of the \two.
This cluster is one of the most stable clusters of
\alli.
A substantial rise in the binding energy
(figure~\ref{fig.energetics}a) is
seen when two Al-Li dimers combine to form \two
~cluster.
Moreover, a peak in the dissociation energy ($\Delta$E)
(figure~\ref{fig.energetics}b), a minima in the second
difference in energy ($\Delta^2$E)
(figure~\ref{fig.energetics}b) and a rather large
HOMO-LUMO gap (figure~\ref{fig.energetics}c) also
signifies a relatively high stability of this cluster.
Interestingly, the planar \two ~structure, with some
bending, has been observed by Kumar~\cite{kumar1} in
the magic cluster Al$_{10}$Li$_8$.
However, because of his restricted nature of search, he
did not find this 8 valence electron system to have
particular stability.

A similar quinted roof structure is also seen in the
ground state structures of the clusters \three ~and
\four (figures~\ref{fig.str1}b(i) and
\ref{fig.str1}c(i)).
However, the \three ~cluster is not energetically as
stable as \two, as seen from
figures~\ref{fig.energetics}a, \ref{fig.energetics}b,
and \ref{fig.energetics}c.
The lowest energy structure of \four ~shows two units
of \two ~(figure~\ref{fig.str1}c(i)).
These two units are arranged so as to form a rectangle
out of the four Al atoms.
Indeed, it is this rectangular shape that gives rise to
an antiaromatic nature in this
cluster~\cite{sharan1,wang2}.
This structure has been extensively studied by
us~\cite{sharan1,sharan3,chacko1} and
others~\cite{wang2}.
The two low-lying structures of \four
~(figure~\ref{fig.str1}c(ii) and \ref{fig.str1}c(iii))
show a square-plane and a quinted roof of four Al
atoms, respectively.

In \five ~cluster, a square-plane of Al$_4$ is seen,
with the fifth Al atom capping this plane forming a
pyramid (figure~\ref{fig.str1}d).
Such a square-planar geometry of Al$_4$ has been
observed in the all-metal aromatic compounds
Al$_4^{2-}$ and $M$Al$_4^-$ ($M$ = Li, Na, and
Cu)~\cite{wang1}.
\five ~shows the first three dimensional structure of
Al$_n$ in \alli ~clusters.
Recall that the appearance of such three dimensional
geometry in pure Al clusters~\cite{rao1} is seen for
Al$_6$.
The \five ~cluster is the most stable among the
clusters studied.
This is clear from the binding energy, $\Delta$E,
$\Delta^2$E, and the HOMO-LUMO gap plots shown in
figures~\ref{fig.energetics}a-\ref{fig.energetics}c,
respectively.
An interesting aspect of the \five ~cluster is the
formation of a plane composed of four Li atoms with one
Al atom at the center, akin to the face of a fcc
cell.
Similar structure for Al$_5$Na$_5$ was predicted by
Dhavale~\cite{dhavale1} \et.
A complete fcc structure can be formed by fourteen
atoms: six Al atoms at the face centered sites and
eight Li atoms at the vertices of the cube.
Such a structure was first studied by Shah~\cite{shah1}
\et. using DBMD.
They found this structure to be the lowest energy
structure of \alsix.
However, our investigation shows that this fcc
structure, with slight distortion, is 1.11eV higher
than that the most stable structure.
The lowest energy structure and the fcc structure of
this cluster is shown in figure~\ref{fig.str1}f(i) and
\ref{fig.str1}f(ii), respectively.
It can be noted that these two structures are nearly
similar, the main difference being the elongation of
one of the Al$_4$ square-planes in the fcc structure
into a rectangle.
It is interesting to note that the series Al$_4^{2-}$,
Li-Al$_4^-$, \five, and \alsix ~shows a three
dimensional octahedral growth of Al cluster surrounded
by Li atoms.
This results in an effective delocalization of the
electron density (see section~\ref{bonding}).

The \six ~cluster also shows a geometry similar to that
of \five, with distortions in the Al$_4$ square-plane
(figure~\ref{fig.str1}e(i)).
In the first low energy structure
(figure~\ref{fig.str1}e(ii)), a tetrahedra of Al$_4$ is
seen, while the other low lying structure
(figure~\ref{fig.str1}e(ii)) shows an octahedra of
Al$_6$.
The octahedral structure is the ground state geometry
of pure Al$_6$ cluster~\cite{rao1}.

The appearance of a pentagonal ring, which is a
precursor to the icosahedral growth, is seen for $n=7$
(Al$_{13}$ is a distorted icosahedron in its ground
state~\cite{ursula1}).
The ground state geometry of \seven ~shows a pentagonal
bipyramid which is capped by the Li atoms
(figure~\ref{fig.str2}a(i)).
Appearance of a similar pentagonal ring in pure Al
clusters (in neutral as well as singly charged form) is
seen for Al$_9$~\cite{rao1} as well as in tetravalent
atom clusters viz.  Sn, and Pb~\cite{bingwang1} for
seven atom clusters.
Incidentally, these clusters are isoelectronic to
\seven ~with 28 valence electrons.
The low energy structure of \seven
~(figure~\ref{fig.str2}a(ii)) as well as the ground
state geometries of $n=8$, 9, and 11 clusters
(figure~\ref{fig.str2}b(i), \ref{fig.str2}c(i) and
\ref{fig.str2}e(i), respectively) also shows such
pentagonal rings.
However, the lowest energy structure of \ten ~cluster
(figure~\ref{fig.str2}d(i)) is completely different
from that of the \seven ~to \nine ~and \eleven, in
that, no signature of the pentagonal ring is seen.
In fact, the geometry is similar to that of Sn$_{10}$,
i.e. a tetracapped trigonal prism (TTP)~\cite{kavita1}
with some distortions.
This TTP structure is enclosed in a Li cage resulting
in its distortion.
It also shows magical behavior, since it exhibits a
peak in the dissociation energy ($\Delta$E), a minima
in the second difference in energy ($\Delta^2$E), and a
large HOMO-LUMO gap
(figure~\ref{fig.energetics}a-\ref{fig.energetics}c,
respectively).

We have performed an analysis of the interatomic bond
distances in order to understand the mixing and
segregation behavior of the Al and the Li atoms.
In figure~\ref{fig.energetics}d, we plot the smallest
bond distances of Al-Al, Al-Li and Li-Li as a function
of number of Al atoms in the \alli ~cluster.
We find that in smaller clusters ($n\le4$) the Al-Al
bond is dominant.
However, from $n=5$ onwards, the Al-Li bond distance is
comparable to that of Al-Al.
Recall that at $n=5$, the first three dimensional Al$_n$
structure emerges, thereby increasing the surface area.
This helps the Li atoms to maximize the number of Al-Li
bonds and thereby move closer to the Al atoms.

It is well known that the density based method is not as
accurate as the Kohn-Sham method, nonetheless, it has
been extensively used to investigate the geometry, and
stability of various Al-Li
clusters~\cite{shah1,shah2,abhijat1}.
Here, we make some pertinent comments on the difference
in the geometry of the \alli ~clusters obtained by
these two methods.
Our results show that the geometries of some of the
clusters were in good agreement with those predicted by
DBMD.
For instance, the structure of \two ~was found to be
a quinted roof by both the methods (see ref.~\ref{shah1}
for DBMD results).
The geometries of other clusters like \seven ~to \nine
~show some substitutional disorder.
However, there are certain clusters where the
geometries were quite different by the KS method.
For instance, the structures of, \three, predicted by
DBMD, is a capped trigonal bipyramid~\cite{shah1}.
When optimized by KS, this structure gets Jahn-Teller
distorted into a structure with a quinted roof of \two.
The evolution in the geometry by DBMD method shows an
earlier appearance of three dimensional structure of
Al$_n$ at $n=4$ (Al$_4$ is tetrahedra in \four), as
compared to the pyramidal structure of Al$_5$ in \five
~by KS method.

\subsection{Bonding}
\label{bonding}

\subsubsection{\two}

As noted above, \two ~is one of the most stable
cluster.
This cluster has 8 valence electrons, which corresponds
to a closed shell in the jellium sense.
However, the behavior of the MOs as well as the
eigenvalue spectrum does not resemble to that of the
SJM.
Kumar~\cite{kumar1} found that the bonding between
Al-Li in the planar \two ~structure is ionic.
On the contrary, our investigation on the bent rhombus
structure does not show signature of any significant
charge transfer from the Li atoms to the Al atoms.
However, an analysis of the difference charge density
shows that, a charge transfer from the Li atoms to the
bonding region between the Al and Li atoms takes place
(figure not shown).
In order to get better insight of the bonding in
\two, we have studied various clusters viz.  Al$_2$,
Al$_2^-$, Al$_2^{2-}$, Al$_2$Li, Al$_2$Li$^-$, and
Al$_2$Li$_2$.
This series represent a growth of \two ~from the Al$_2$
dimer.
In figure~\ref{fig.al2}, we show the eigenvalue
spectrum, and the Al-Al bond distance for the optimized
geometries of these clusters.
The eigenvalue spectrum of Al$_2$
(figure~\ref{fig.al2}a), in neutral, singly and doubly
charged form, shows a triply degenerate HOMO state.
These states are partially filled two $\pi$ and one
$\sigma$ bond (figure not shown).
Successive substitution of each electron in the charged
dimer, by Li atoms splits the HOMO state reducing the
degeneracy.
The first Li atom forms a $\sigma$ bond with the
$p_y-p_y ~ \pi$ orbital of the Al$_2$ dimer (isodensity
surfaces not shown).
The second Li atom further splits the HOMO state
introducing a substantial gap of 0.68eV.
However, in this case, a hybridization of the $s$
orbital of Li atom and the $\pi$ orbital of Al$_2$
takes place.
In figure~\ref{fig.al2.mo}, we show the second HOMO
and the HOMO state of the \two ~cluster.
Clearly, these states shows the $\pi$ bonded orbitals
of Al$_2$ hybridized with the $s$ orbitals of the Li
atoms.
In figure~\ref{fig.al2}b, we plot the Al-Al bond
distance in these clusters.
A contraction of the Al-Al bond upon addition of an
electron or a Li atom is seen.
The contraction of the bond on addition of a Li atom
being more than that due to the addition of an
electron.
Thus, the hybridization of the $s$ orbitals of Li with
the $\pi$ orbitals of Al$_2$ enhances its stability.

\subsubsection{\four}

The bonding in \four ~cluster has been extensively
studied by us~\cite{sharan1,sharan3,chacko1} and
others~\cite{wang2}.
This cluster has been found to be antiaromatic with
four $\pi$-electrons.
For the sake of completeness, we discuss the bonding in
this cluster.
This cluster, as discussed earlier, is composed of two
\two ~units arranged edge-to-edge to form a Al$_4$
rectangle.
This rectangular structure leads to antiaromaticity.
In this cluster $sp^2$ hybridization of Al takes place
leaving one empty unhybridized $p$ orbital.
The valence electron of each of the four Li atoms is
then transfered to this empty $p_z$ orbital, thus
providing four $\pi$-electrons for antiaromaticity.
Interestingly, its first low-lying structure
(figure~\ref{fig.str1}c(ii)) can also be considered as
a candidate for antiaromaticity.
An analysis of the MOs shows that this cluster also
has four $\pi$-electrons similar to that of the lowest
energy structure, thereby conforming the structural and
electron count criteria for antiaromaticity.
However, a detailed analysis of the magnetic field
induced ring currents has to be done in order to
understand the antiaromaticity in this structure.

\subsubsection{\five ~and \alsix}

As discussed in section \ref{structure}, the structures
of \five ~and \alsix ~(in fcc geometry) clusters show a
square-plane of Al$_4$.
Such Al$_4$ square-plane in Al$_4^{2-}$ and Li-Al$_4^-$
has lead to aromaticity in these clusters~\cite{wang1}.
Moreover, these clusters (Al$_4^{2-}$, Li-Al$_4^-$,
\five, and \alsix) show a three dimensional octahedral
growth of the Al cluster.
It is, therefore, interesting to discuss the change in
bonding, and hence the effect on aromaticity in
these clusters.
Recall that Al$_4^{2-}$ and Li-Al$_4^-$ are aromatic
clusters with two $\pi$-electrons~\cite{wang1}.
The HOMO state in these clusters are completely
delocalized $\pi$ orbitals.
There are two more delocalized $\sigma$ bonds: one
composed of radial $p$ orbitals, and the other composed
of perpendicular $p$ orbitals.
Due to the presence of a Al$_4$ square-plane, the
bonding in \five ~and \alsix ~clusters is expected to
be similar to that in Al$_4^{2-}$.
In that case, these clusters would have
two $\pi$-electrons similar to that of Al$_4^{2-}$ and
Li-Al$_4^-$, and show some signature of aromaticity.
In figure~\ref{fig.al-square}a and
\ref{fig.al-square}b, we show the HOMO states of \five
~and \alsix.
Interestingly, the HOMO state of \five ~does show such
a delocalized $\pi$ bond between the Al$_4$
square~(figure~\ref{fig.al-square}a).
However, a lone electron is seen at the fifth
Al atom that caps this square.
This reduces the number of $\pi$ electrons count to
one, thereby losing the aromaticity.
The HOMO state of \alsix, on the other hand, is quite
different than that seen in the previous clusters
(see figure~\ref{fig.al-square}b).
It shows a localized bond composed of $p$ orbitals of
the Al atoms.
In figure~\ref{fig.evs}, we show the eigenvalue
spectrum of \alsix, along with that of the other
\alli ~($n=$1-11) and Sn$_{10}$ clusters.
It is clear from this that the eigenvalue spectrum of
\alsix ~shows nearly a jellium-like structure.
However, an analysis of the behavior of the molecular
orbitals (figure not shown) of \alsix ~shows that the
lower six states are jellium-like, whereas, the higher
states show localized bonds arising out of $p$ orbitals
of the Al atoms.
The isodensity surfaces for the 6$^{\rm th}$ state of
\five ~and the 10$^{\rm th}$ state of \alsix ~are shown
in figure~\ref{fig.al-square}c, and
\ref{fig.al-square}d, respectively.
These are $\sigma$-bonded states, composed of radial
$p$ orbitals and are analogous to the radial $p$
orbitals of Al$_4^{2-}$, as noted by Wang
\et~\cite{wang1}.
Since, these bonds involve all the Al atoms, they
exhibit a substantial delocalization.

\subsubsection{\alli ~($n$=7-11)}

The bonding in these clusters (\alli, $n=7-11$), is
quite different from that of the smaller clusters.
The analysis of the behavior of the MOs shows that the
eigenvalue spectrum can be classified into
two groups: the lower $n$ states are jellium-like,
whereas, the higher states forms a complex band of
localized bonds arising out of the interaction of the
$p$ electrons of each of the Al atoms.
The lower states can be identified as {\em 1s, 1p, 1d,
2s,...}, in conformation to the SJM.
Typical representative isodensity plots of these
states are shown in figure~\ref{fig.jellium}.
They represent the {\em 1s, 1p, 1d} and {\em 2s}
states of the jellium composed of the Al-$3s$ orbitals.
In figure~\ref{fig.evs}, we show the eigenvalue
spectrum of \alli ~($n=$1-11) and Sn$_{10}$ clusters.
This figure shows a clear energy separation of lower
jellium-like $n$ states of the clusters with $n>6$,
from the higher orbitals.
In order to study this more extensively, we have
performed a spherical harmonics analysis of the KS
orbitals~\cite{spchar1}.
We find that the lower $n$ orbitals have a predominant
$s$~character (75-85\%) indicating that these state are
arising out of the $3s$ orbitals of the Al atoms,
which hybridizes to form a jellium.
However, this description cannot be extended to the
higher states.
The higher states are composed of Al-$3p$ orbitals.

The localization characteristics in the bonding in the
Al-Li clusters is also analyzed via the electron
localization function plots.
Previous investigations show that even in large Al-Li
clusters, the Al-Al bond is covalent.
However, our earlier work on the structural properties
shows a clustering of Al atoms enclosed in Li cage.
This clustering could lead to a delocalization of the
electron density.
In order to study this, we have analyzed the electron
localization function in the \alli ~clusters.
In figure~\ref{fig.elf}, we show the isovalued surface
of the ELF for the clusters \two, \five, and \ten,
at the values 0.85, 0.8, and 0.72, respectively.
These plots shows a localization of the electron
density along the Al-Al bond (increasing the value of
ELFs does not show any bond between Al-Al).
It has been noted by Silvi \et~\cite{silvi1}, that an
ELF value of about 0.7 or greater is a indication of a
localized bond in that region.
Thus, the Al-Al bonds in these clusters are
predominantly covalent, in accordance with the previous
observation.
However, as the cluster grows in size, the degree of
localization of the electron density in these clusters
is reduced.
Further, the localization in \five ~is only in the
plane forming the square (for out of plane, the Al-Al
bond is seen for the ELF value of 0.7).
A similar behavior of the ELF for \seven ~(not shown in
figure) is seen, in that, a localized bond along the
pentagon (ELF value = 0.78) and delocalized bond in
the perpendicular direction (ELF value = 0.6) is seen.

\def \al{Al }
\section {Tetravalent behavior of \al in \alli ~clusters}
\label{tetra}

Previous investigations on mixed Al-Li
clusters~\cite{kumar1,gong1,gong2}, have shown a
substantial charge transfer from the Li atom to the Al
atom.
Due to such charge transfer, Al behaves as a tetravalent
atom in mixed Al-Li clusters with nearly 50:50 percent
concentration.
In this section, we present the evidence of such
behavior.
The analysis of the geometries of Al$_n$ in \alli
~shows remarkable similarities to those of the clusters
of the tetravalent atoms viz., Si, Ge, Sn and Pb.
The transition from planar to three dimensional
geometry at $n=5$, the appearance of a pentagonal ring
at $n=7$, and the formation of the distorted TTP
structure of 10 atom cluster are some of the features
of the clusters of tetravalent atoms.
Such behavior of Al$_n$ in \alli ~clusters is also
observed.
Moreover, the eigenvalue spectrum, in
figure~\ref{fig.evs}, of the \ten ~show remarkable
similarity to that of the Sn$_{10}$ cluster.
The bonding between Al-Al in these clusters is
covalent, which is similar to that in the clusters of
tetravalent atoms.
Finally, the resemblance of the DOS of the Al-Li in its
most stable bulk phase i.e. B32, to that of
diamond~\cite{das1} also indicates a tetravalent
behavior of the Al atom in Al-Li systems with 50:50
concentration.

\section {Conclusion}
\label{conclusion}
In the present work, we have reported the systematic
investigation of the geometries, energetics, electronic
structure and bonding in various Al-Li clusters
viz. \alli ~($n$=1-11), Al$_2^-$, Al$_2^{2-}$,
Al$_2$Li, Al$_2$Li$^-$, and \alsix ~using BOMD within
the framework of density functional theory.
The bonding in these clusters was discussed via the
electron localization function as well as the behavior
of the molecular orbitals.
The growth pattern is divided into two broad
categories: first consisting of a bent rhombus of \two
~($n$=2-4) and second consisting of a pentagonal ring
($n$=7-9,11).
We find that the 8, 20, and 40 valence electron systems
to be magic, exhibiting a peak in the dissociation
energy, a minima in the second difference in energy,
and a large HOMO-LUMO gap.
The structural transition of Al$_n$ in \alli, from two
dimension to three dimension increases the surface
area, thereby helping the Li atoms to maximize the
number of Al-Li bonds, as a result of which the Li
atoms move closer to the Al atoms.
In clusters with more than six Al atoms, a charge
transfer from Li atoms to Al atoms makes Al behave as a
tetravalent atom like Si, Ge, Sn, and Pb in \alli
~clusters.
These negatively charged Al$_n$ structure is then
stabilized by the positive Li environment.
The behavior of the MOs in these clusters ($n>6$) can
be divided two groups: a lower group of jellium-like
states arising out of the $s$ electrons of the Al atoms
and a higher group of localized bonds formed through
the interaction of the $p$ electrons on each of the Al
atoms.
The formation of the three dimensional Al$_n$ structure
destroys the aromatic and antiaromatic nature in these
clusters.

\section {Acknowledgement}
\label{ack}

SC gratefully acknowledges the financial support of CSIR
(New Delhi).
VVP wishes to thank NSERC of Canada and the University
of Pune for partially supporting this research.

\newpage

{\Large List of Figures}

\begin{center}
\begin{figure}
        \epsfxsize=4.0in \epsffile{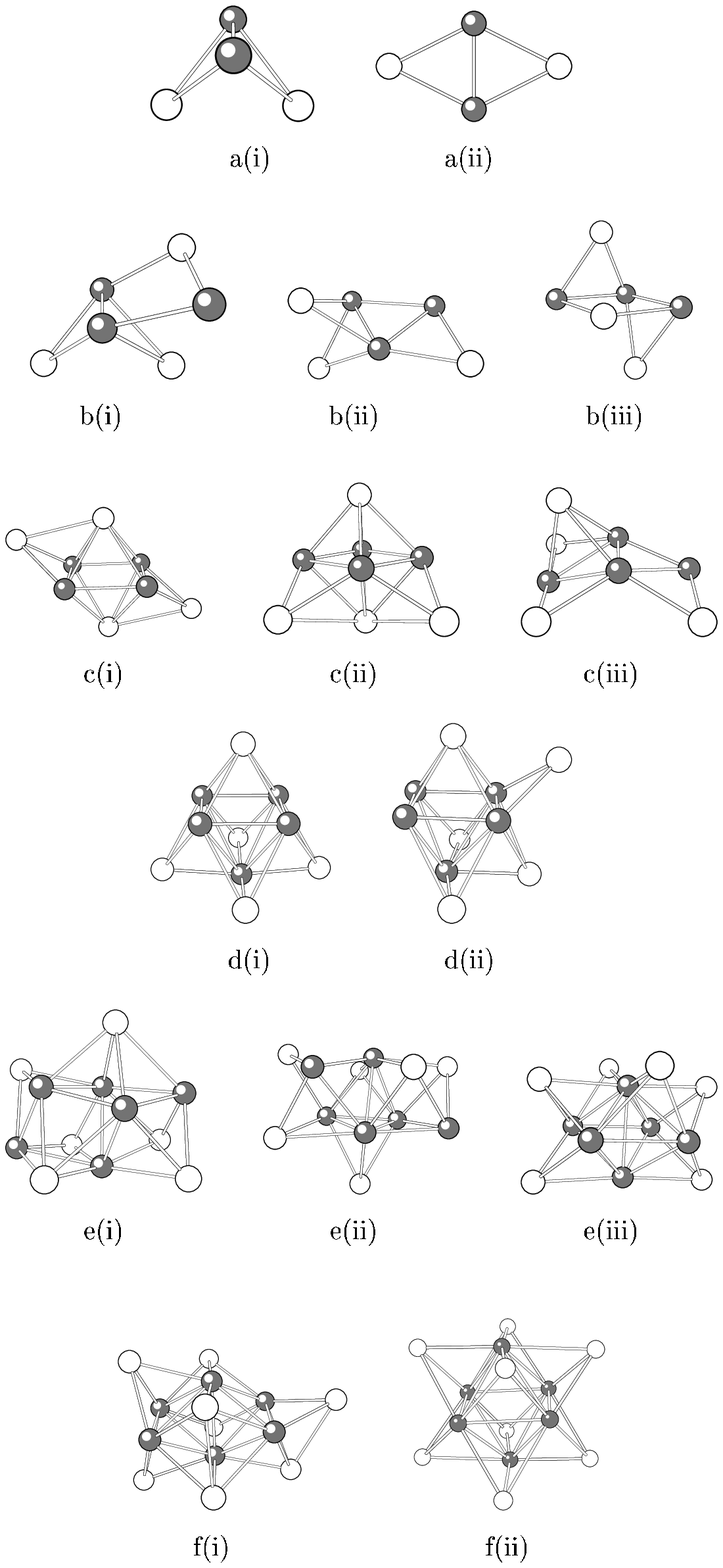}
        \caption{}
        \label{fig.str1}
\end{figure}
\end{center}

\begin{center}
\begin{figure}
        \epsfxsize=4.0in \epsffile{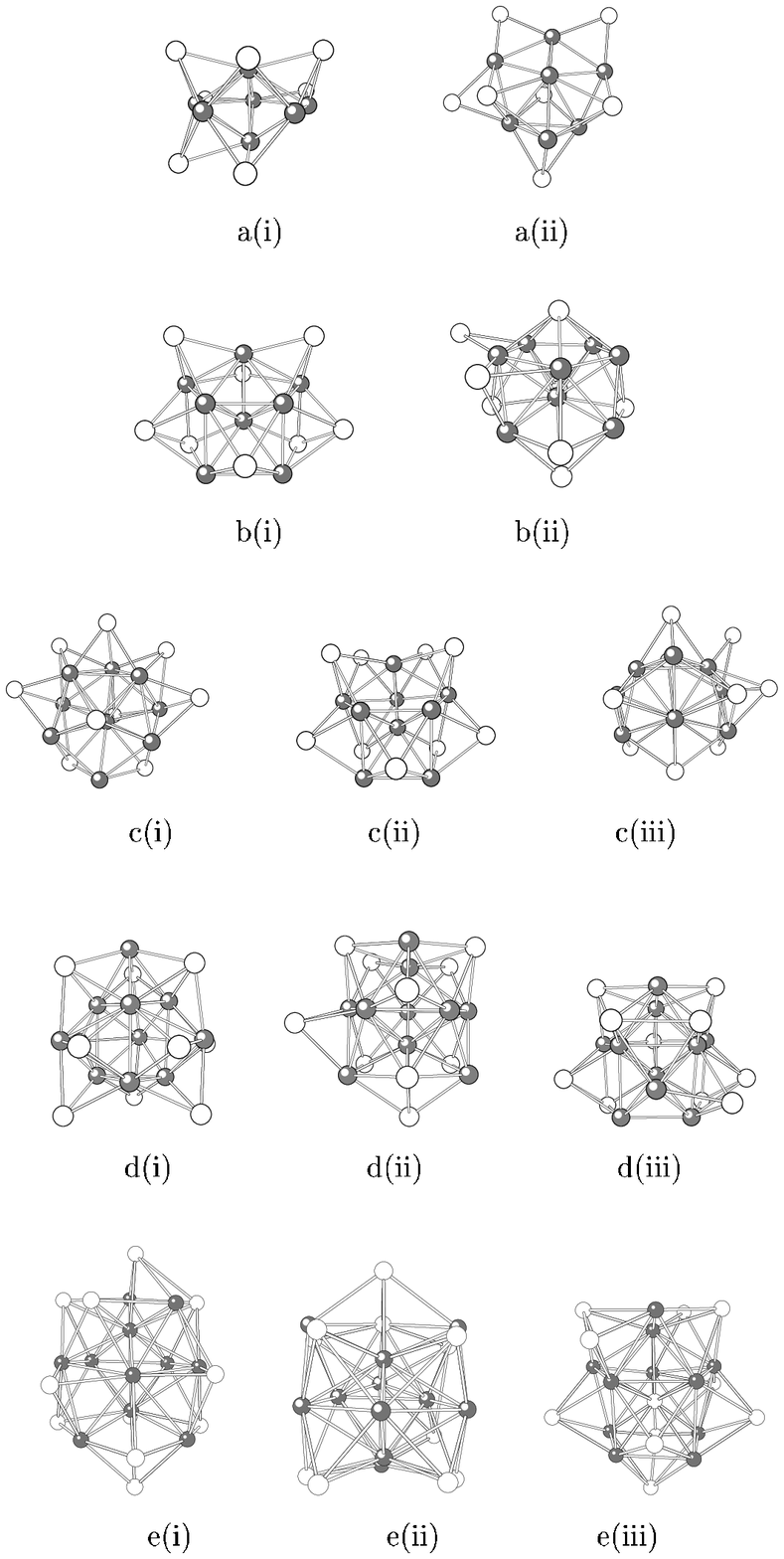}
        \caption{}
        \label{fig.str2}
\end{figure}
\end{center}

\begin{center}
\begin{figure}
        \epsfxsize=5.0in \epsffile{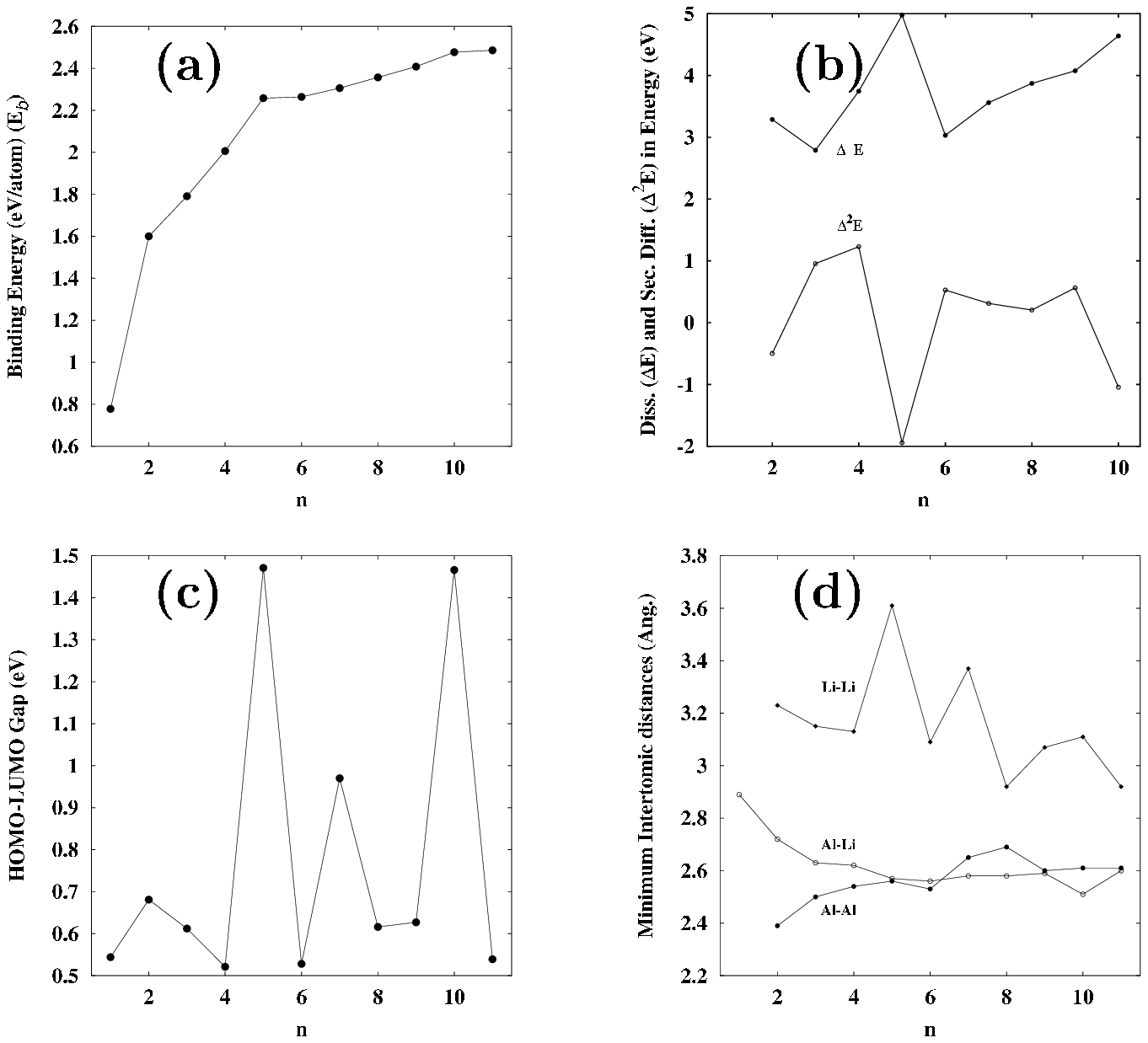}
        \caption{}
        \label{fig.energetics}
\end{figure}
\end{center}

\begin{center}
\begin{figure}
        \epsfxsize=4.0in \epsffile{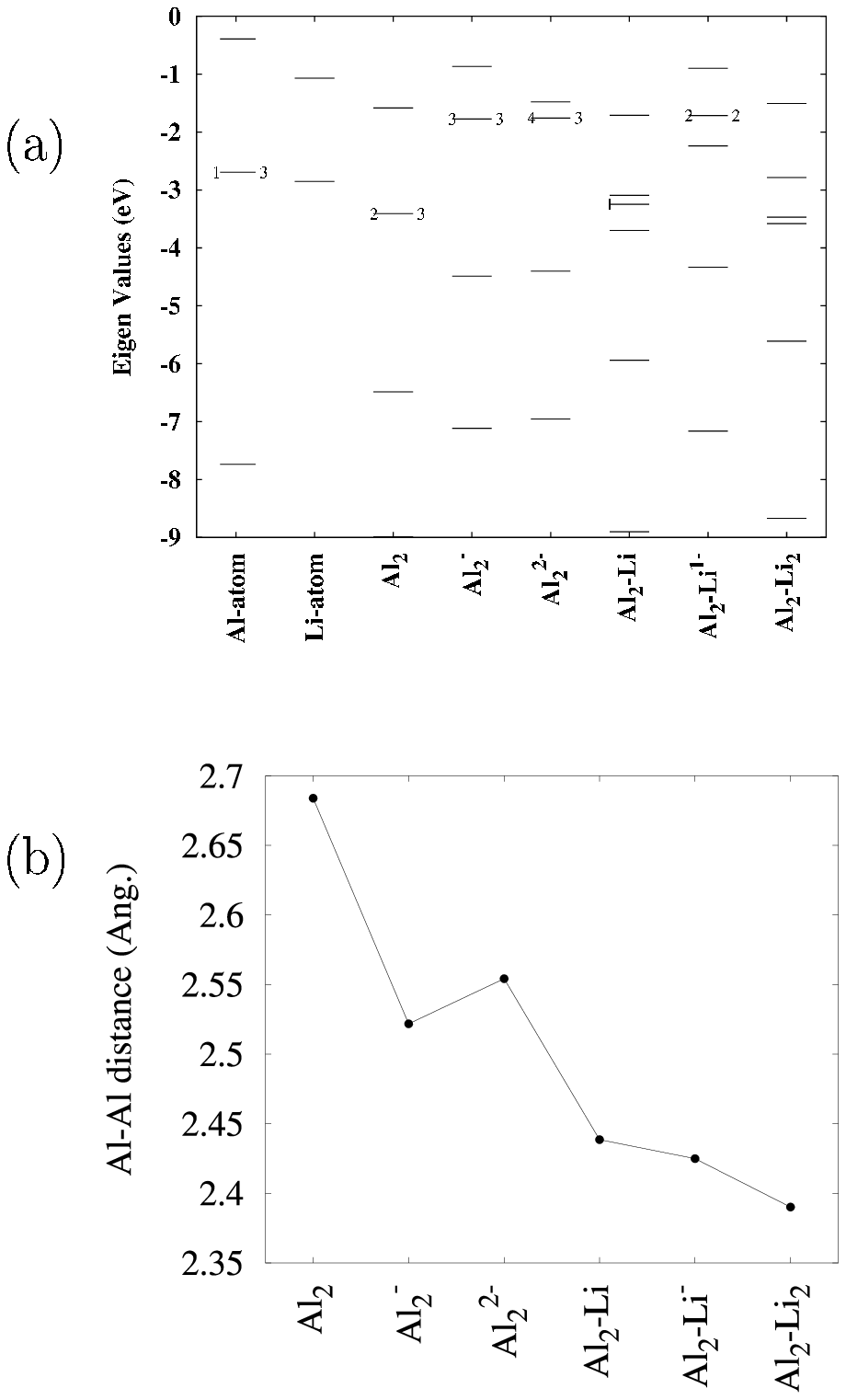}
        \caption{}
        \label{fig.al2}
\end{figure}
\end{center}

\begin{center}
\begin{figure}
        \epsfxsize=5.0in \epsffile{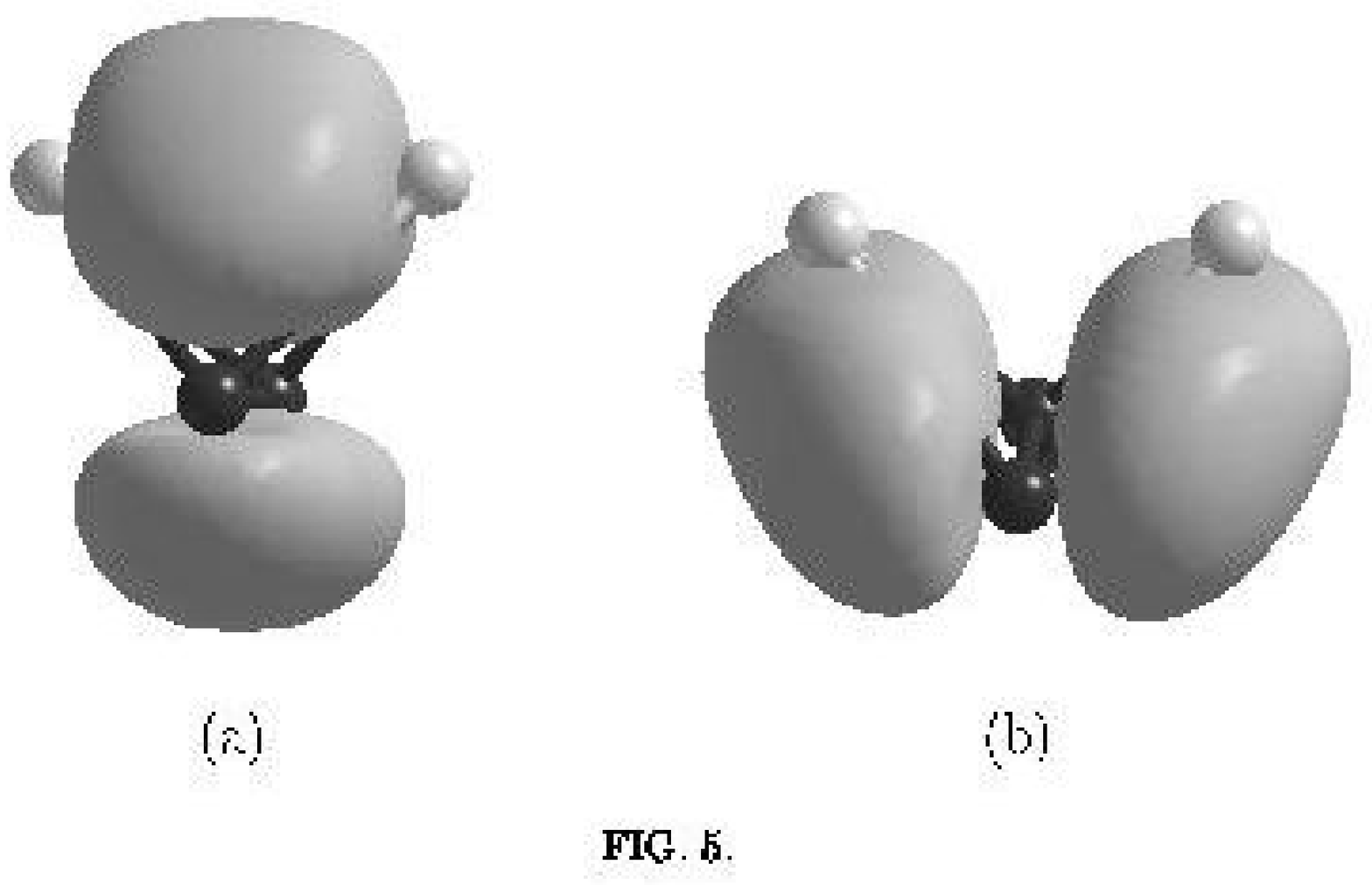}
        \caption{}
        \label{fig.al2.mo}
\end{figure}
\end{center}

\begin{center}
\begin{figure}
        \epsfxsize=5.0in \epsffile{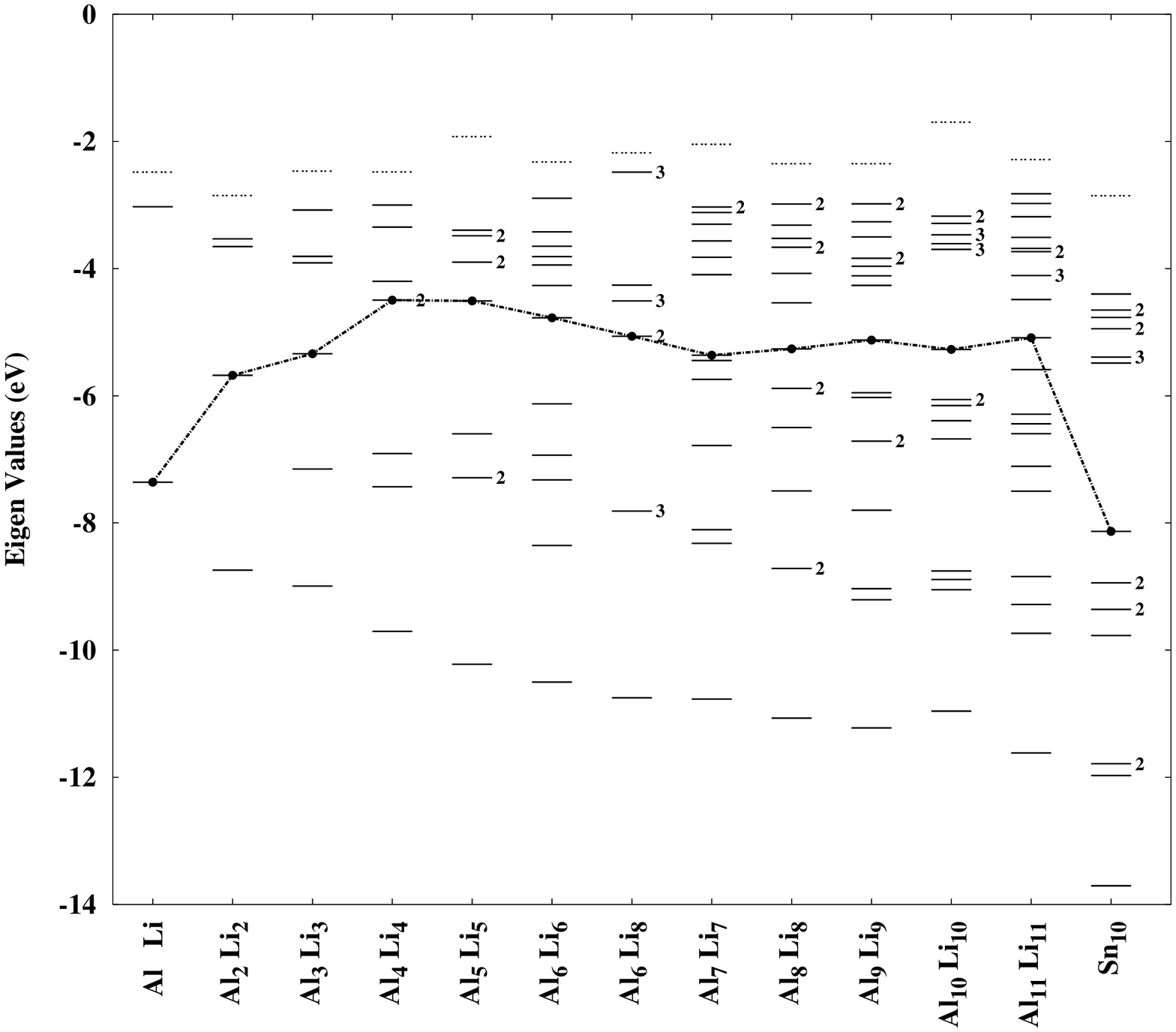}
        \caption{}
        \label{fig.evs}
\end{figure}
\end{center}

\begin{center}
\begin{figure}
        \epsfxsize=5.0in \epsffile{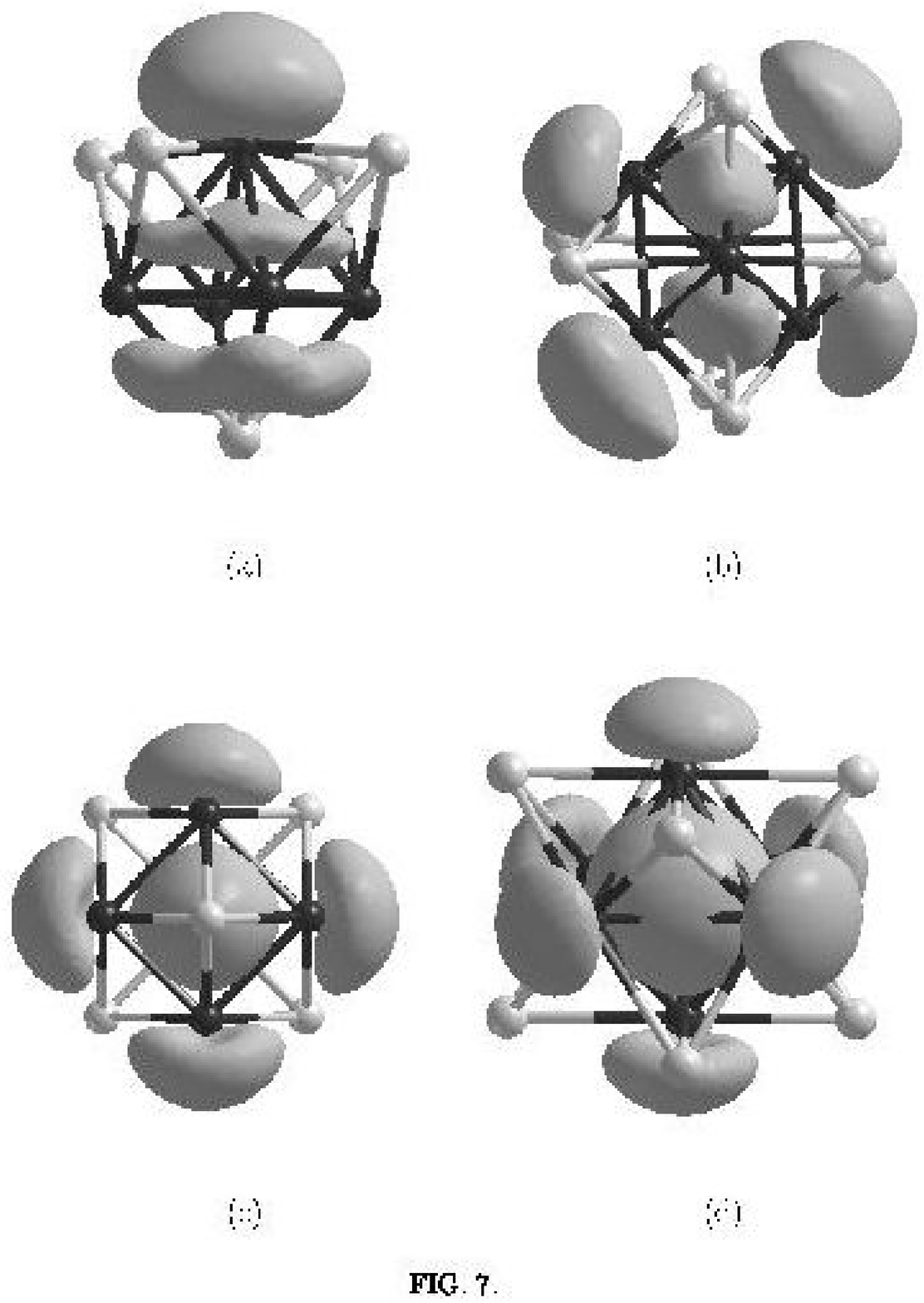}
        \caption{}
        \label{fig.al-square}
\end{figure}
\end{center}

\begin{center}
\begin{figure}
        \epsfxsize=5.0in \epsffile{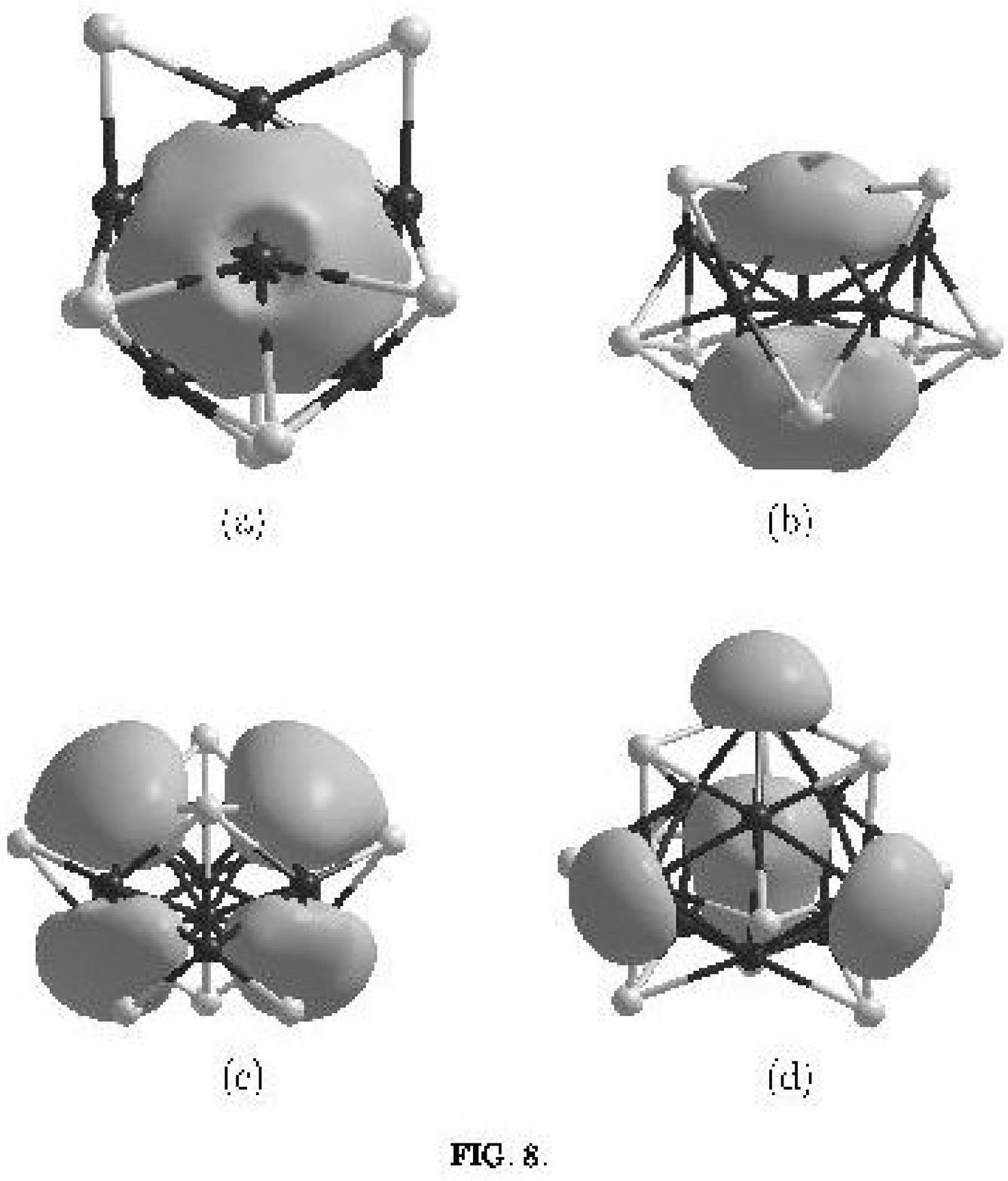}
        \caption{}
        \label{fig.jellium}
\end{figure}
\end{center}

\begin{center}
\begin{figure}
        \epsfxsize=5.0in \epsffile{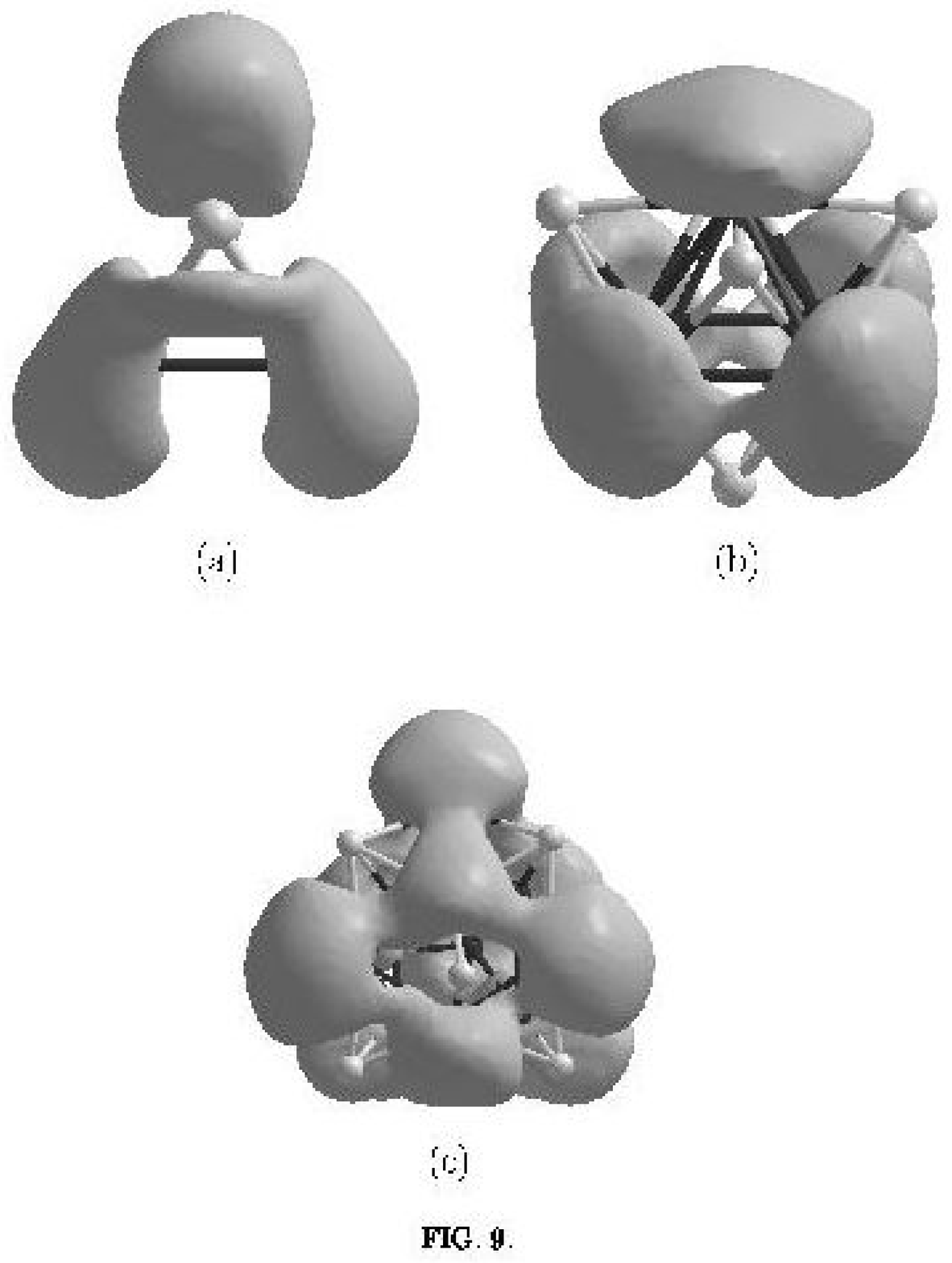}
        \caption{}
        \label{fig.elf}
\end{figure}
\end{center}

\begin{enumerate}
  \item [ \ref{fig.str1} ]
         The ground state and the low-lying geometries
         of the \alli ~($n$=2-6) and \alsix ~clusters.
         The black circles represent Al atoms and the
         white circles represent the Li atoms.
         The lowest energy structure is represented
         by~(i).

  \item [ \ref{fig.str2} ]
         The ground state and the low-lying geometries
         of the \alli ~($n$=7-11) clusters.
         The black circle represents Al atoms and the
         white circles represent the Li atoms.
         The lowest energy structure is represented
         by~(i).

  \item [ \ref{fig.energetics} ]
         (a)~The binding energy per atom (in eV/atom)
             of the \alli ~cluster ($n$=1-11).
         (b)~The dissociation energy ($\Delta$E)
             and the second difference in the total
             energy ($\Delta^2$E) of the \alli
             ~clusters ($n$=2-10) with respect to a
             Al-Li dimer.
         (c)~The HOMO-LUMO gap (in eV) of the \alli
             ~cluster ($n$=1-11).
         (d)~The minimum interatomic distance Al-Al,
             Al-Li and Li-Li (in \AA) of the \alli
             ~cluster ($n$=1-11).

  \item [ \ref{fig.al2} ]
         (a)~The eigenvalue spectrum (in eV).
             The continuous lines correspond to
             occupied states and dashed lines
             correspond to the empty states.
             The numbers on the right indicate
             the degeneracy of the states.
             The numbers on the left indicate
             the occupancies of the states.
             All other occupied states have
             two electrons.
         (b)~The Al-Al bond distance (in \AA).

  \item [ \ref{fig.al2.mo} ]
         The isodensity surface of the second HOMO and
         the HOMO state of \two ~at 1/5$^{\rm th}$ of
         its maximum value.
         The black circles represent Al atoms and the
         gray circles represent the Li atoms.

  \item [ \ref{fig.evs} ]
         The eigenvalue spectrum (in eV) of \alli
         ~($n$=1-11), \alsix, and Sn$_{10}$ clusters.
         The continuous lines correspond to the
         occupied states and the dotted lines
         correspond to the empty states.
         The numbers on right indicate the
         degeneracy of the states.
         All states are doubly occupied.
         The thick dashed line shows the $n^{\rm
         th}$ state of these clusters, where $n$
         is the number of Al atoms.
         For Sn$_{10}$ the thick dashed line
         corresponds to the 10$^{\rm th}$ state.

  \item [ \ref{fig.al-square} ]
         (a)~The isodensity surface of the HOMO state
             of \five, at 1/5$^{\rm th}$ of its
             maximum.
         (b)~The isodensity surface of the HOMO state
             of \alsix, at 1/5$^{\rm th}$ of its
             maximum.
         (c)~The isodensity surface of the $6^{\rm th}$
             state of \five, at 1/5$^{\rm th}$ of its
             maximum.
         (d)~The isodensity surface of the $10^{\rm
             th}$ state of \alsix, at 1/8$^{\rm th}$ of
             its maximum.
         In all the figures, the black circles
         represent Al atoms and the gray circles
         represent the Li atoms.

  \item [ \ref{fig.jellium} ]
         The isodensity surface of various orbitals.
         These are:
            (a)~1$^{\rm st}$ orbital of \seven ~at
                1/5$^{\rm th}$ of its maximum value,
            (b)~3$^{\rm rd}$ orbital of \eight ~at
                1/5$^{\rm th}$ of its maximum value,
            (c)~8$^{\rm th}$ orbital of \nine ~at
                1/10$^{\rm th}$ of its maximum value,
                and
            (d)~10$^{\rm th}$ orbital of \ten ~at
                1/5$^{\rm th}$ of its maximum value,
         representing the {\em 1s, 1p, 1d}, and {\em 2s}
         state of the jellium composed of the Al-3$s$
         orbital.
         In all the figures, the black circles
         represent Al atoms and the gray circles
         represent the Li atoms.

  \item [ \ref{fig.elf} ]
         The isodensity surface of the electron
            localization function (ELF) of:
            (a)~\two ~at the value 0.85,
            (b)~\five ~at the value 0.80, and
            (c)~\ten ~at the value 0.72.
         In all the figures, the black circles
         represent Al atoms and the gray circles
         represent the Li atoms.

\end{enumerate}

\newpage
\begin{figure}
\end{figure}

\newpage

\newpage

\newpage

\newpage

\newpage

\newpage

\newpage

\newpage

\newpage

\begin{references}
\bibitem{spchar1}
  \label{spchar1}
  The character in an orbitals is calculated by
    projecting the orbitals onto spherical harmonics
    centered at each ionic sites within a sphere of a
    specified radius around each ion.
  The radius of the sphere is taken to be half of the
    distance of the ion from the nearest ion.
\end{references}
\end{document}